\documentstyle[epsf,preprint,tighten,prd,aps]{revtex}

\newcommand{\Lie}{{\pounds}}
\renewcommand{\Re}{\text{Re} }

\begin{document}
\epsfverbosetrue
\draft

%%% Front matter

%\wideabs{%%% Remove before submission!
\title{
  Critical Collapse Beyond Spherical Symmetry:\\
  General Perturbations of the Roberts Solution
}
\author{Andrei V. Frolov%
  \thanks{Email: \texttt{andrei@phys.ualberta.ca}}}
\address{
  Physics Department, University of Alberta\\
  Edmonton, Alberta, Canada, T6G 2J1}
\date{\today}
\maketitle

\begin{abstract}
  This paper studies the non-spherical perturbations of the
  continuously self-similar critical solution of the gravitational
  collapse of a massless scalar field (the Roberts solution). The exact
  analysis of the perturbation equations reveals that there are no
  growing non-spherical perturbation modes.
\end{abstract}

%}
\narrowtext

%%% Main body

\section{Introduction} \label{sec:intro}

Choptuik's discovery of critical phenomena in the gravitational collapse of
a scalar field \cite{Choptuik:93} sparked a surge of interest in
gravitational collapse just at the threshold of black hole formation.
The discovery of critical behavior in several other matter models quickly
followed \cite{Abrahams&Evans:93,Evans&Coleman:94,Koike&Hara&Adachi:95,%
Maison:96,Hirschmann&Eardley:95a,Hirschmann&Eardley:95b}. Despite the
fact that the evolution equations are very complex and highly
non-linear, the dynamics of the near-critical field evolution is relatively
simple and, in some important aspects, universal. The critical solution,
which depends on the matter model only, serves as an intermediate
attractor in the phase space of solutions, and often has an additional
peculiar symmetry called self-similarity. The mass of the black hole
produced in supercritical evolution scales as a power law
\begin{equation} \label{eq:scaling}
  M_{\text{BH}}(p) \propto |p-p^*|^\beta,
\end{equation}
with parameter $p$ describing initial data, and mass-scaling exponent
$\beta$ is dependent only on the matter model, but not on the initial
data family. An interesting consequence of mass scaling which has
direct bearing on the cosmic censorship conjecture is the fact that
arbitrarily small black holes can be produced in near-critical
collapse, with the critical solution exhibiting a curvature singularity
and no event horizon.

The explanation of the universality of the critical behavior lies in
perturbation analysis and renormalization group ideas 
\cite{Evans&Coleman:94,Koike&Hara&Adachi:95,Maison:96,Hara&Koike&Adachi:96}.
It turns out that critical solutions generally have only one unstable
perturbation mode, making them the most important solutions for
understanding the dynamics of field evolution, after the stable ones
(flat space and Schwarzschild or Kerr-Newman black hole). As the
near-critical field configuration evolves, all its perturbation modes
decay, losing information about the initial data and bringing the
solution closer to critical, except the one growing mode which will
eventually drive the solution to black hole formation or dispersal,
depending on its content in the initial data. Thus the critical
solution acts as an intermediate attractor (of codimension one) in the
phase space of field configurations. Finding the eigenvalue of the
growing perturbation mode allows one to calculate important parameters
of the critical evolution, the mass-scaling exponent in particular.

An important question is how generic the critical behavior is with
respect to initial data, or, in phase space language, how big is the
basin of attraction of the critical solution. So far most of the work
on critical gravitational collapse, numerical or analytic, has been
restricted to the case of spherical symmetry, simply because of the
enormous difficulties in treating fully general non-symmetric solutions
of Einstein equations. A natural concern is whether the critical
phenomena observed so far are limited to spherical symmetry, and
whether the evolution of non-spherical data will lead to the same
results. The numerical study of Abrahams and Evans on axisymmetric
gravitational wave collapse \cite{Abrahams&Evans:93} and recent
numerical perturbation calculations by Gundlach
\cite{Gundlach:97,Gundlach:98} give numerical evidence for the claim
that critical phenomena are not restricted to spherical symmetry, and
that the critical solutions are indeed attractors in the full phase
space. In this paper we search for analytical evidence to support that
claim.

One of the few known closed form solutions related to critical
phenomena is the Roberts solution, originally constructed as a
counterexample to the cosmic censorship conjecture \cite{Roberts:89},
and later rediscovered in the context of critical gravitational
collapse \cite{Brady:94,Oshiro&Nakamura&Tomimatsu:94}. It is a
continuously self-similar solution of a spherically symmetric
gravitational collapse of a minimally coupled massless scalar field.
While it is not a proper critical solution, as it has more than one
growing mode \cite{Frolov:97}, it is still a good (and simple) toy
model of the critical collapse of the scalar field.

This paper considers fully general perturbations of the Roberts
solution in a gauge-invariant formalism. Due to the symmetries of the
background, the linear perturbation equations decouple and the
variables separate, so an exact analytical treatment is possible. We
find that there are no growing perturbation modes apart from
spherically symmetric ones described earlier \cite{Frolov:97}. So all
the non-sphericity of the initial data decays in the collapse of the
scalar field, and only the spherically symmetric part will play a role
in the critical behavior.

To our knowledge, this is the first paper to obtain analytical results
on non-spherical critical collapse.

\section{The Roberts Solution} \label{sec:roberts}

\begin{figure}
  \begin{center}\begin{tabular}{cc}
    \epsfxsize=\textwidth \multiply\epsfxsize 4 \divide\epsfxsize 10
    \epsfbox{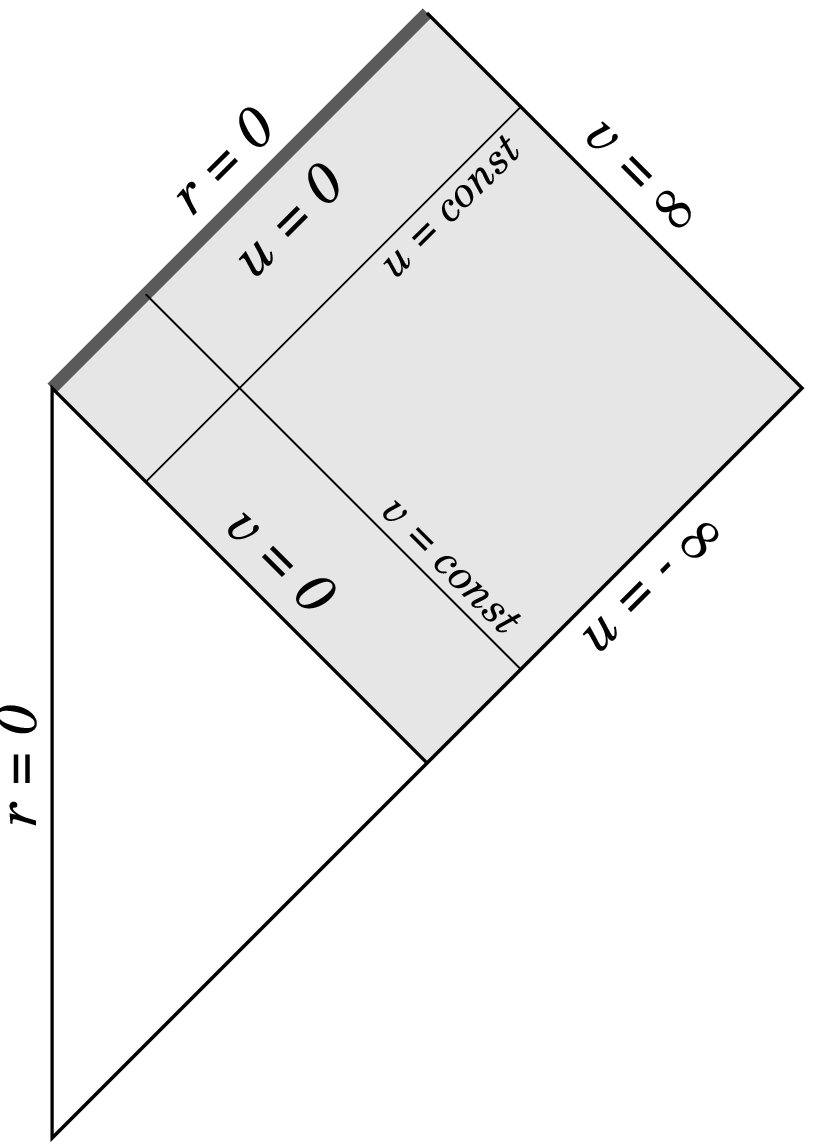} &
    \epsfxsize=\textwidth \multiply\epsfxsize 4 \divide\epsfxsize 10
    \epsfbox{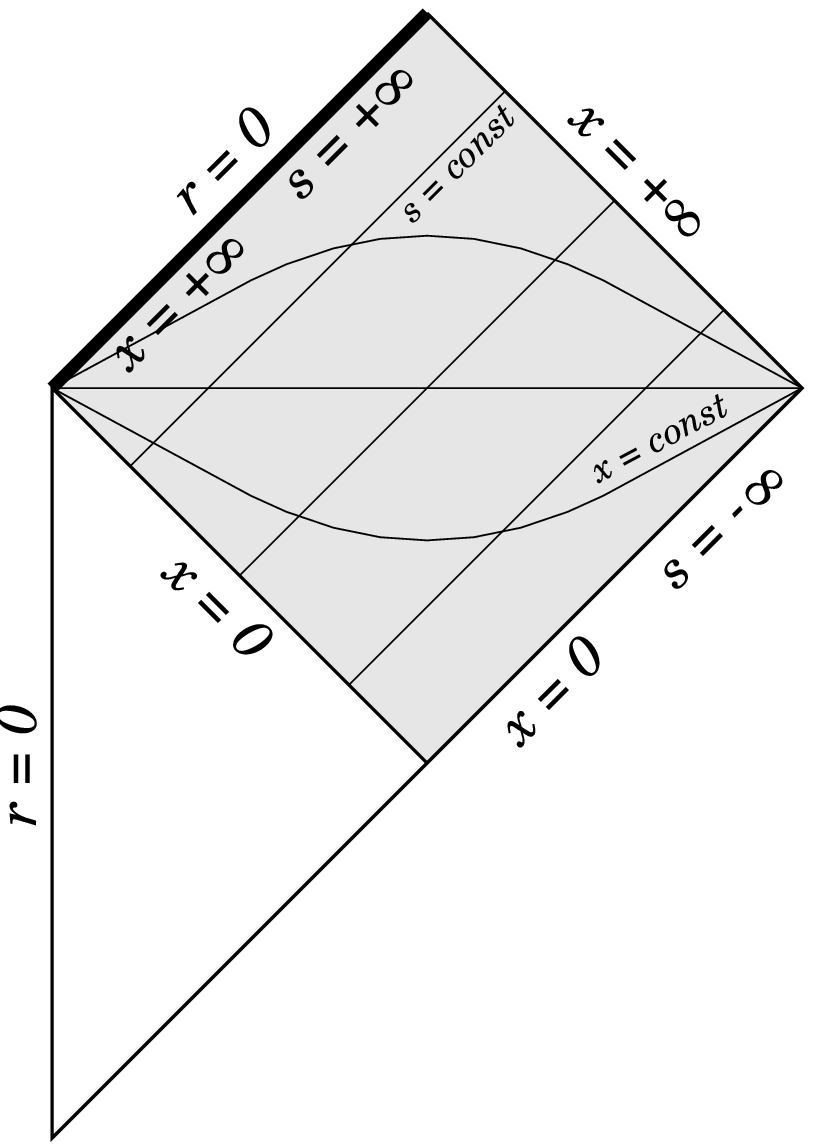}\\
    (null coordinates) & (scaling coordinates)
  \end{tabular}\end{center}
  \caption{
    Global structure of the Roberts solution: The scalar field influx
    is turned on at $v=0$; spacetime is flat before that. Field
    evolution occurs in the shaded region of the diagram, and there is
    a null singularity in the center of the spacetime.
  } 
  \label{fig:roberts}
\end{figure}

The spacetime we will use as a background in our calculations is a
continuously self-similar spherically symmetric solution of the
gravitational collapse of a massless scalar field (the Roberts
solution). The Einstein-scalar field equations
\begin{eqnarray}
  R_{\mu\nu} &=& 2 \phi_{,\mu} \phi_{,\nu}, \label{eq:r}\\
  \Box\phi &=& 0 \label{eq:box}
\end{eqnarray}
can be solved analytically in spherical symmetry by imposing continuous
self-similarity on the solution, i.e. by assuming that there exists a
vector field $\xi$ such that
\begin{equation} \label{eq:ss}
  \Lie_\xi g_{\mu\nu} = 2g_{\mu\nu}, \hspace{1em}
  \Lie_\xi \phi = 0,
\end{equation}
where $\Lie$ denotes Lie derivative. Self-similar solutions form a
one-parameter family, which is most easily derived in null coordinates
\cite{Brady:94,Oshiro&Nakamura&Tomimatsu:94,Frolov:98}. The critical
solution is given by the metric
\begin{equation} \label{eq:metric}
  ds^2 = - 2\, du\, dv + r^2\, d\Omega^2,
\end{equation}
where
\begin{equation} \label{eq:crit}
   r = \sqrt{u^2 - uv}, \hspace{1em}
  \phi = \frac{1}{2} \ln \left[1 - \frac{v}{u}\right].
\end{equation}
The global structure of the critical spacetime is shown in
Fig.~\ref{fig:roberts}. The influx of the scalar field is turned on at
the advanced time $v=0$, so that the spacetime is Minkowskian to the
past of this surface. The initial conditions for the field equations
(\ref{eq:r}) and (\ref{eq:box}) are specified there by the continuity
of the solution.

It is instructive to rewrite Roberts solution in new coordinates so
that the self-similarity becomes apparent. For this purpose we
introduce scaling coordinates
\begin{equation} \label{eq:coord:xs}
  x = \frac{1}{2} \ln \left[1 - \frac{v}{u}\right], \hspace{1em}
  s = - \ln(-u),
\end{equation}
with the inverse transformation
\begin{equation} \label{eq:coord:uv}
  u = - e^{-s}, \hspace{1em}
  v = e^{-s} (e^{2x} - 1).
\end{equation}
The signs are chosen to make the arguments of the logarithm positive in
the region of interest ($v>0$, $u<0$), where the field evolution
occurs. In these coordinates the metric (\ref{eq:metric}) becomes
\begin{equation} \label{eq:metric:xs}
  ds^2 = 2 e^{2(x - s)}
    \left[(1 - e^{-2x}) ds^{2} - 2 ds dx\right] +
    r^2\, d\Omega^2,
\end{equation}
and the critical solution (\ref{eq:crit}) is simply
\begin{equation} \label{eq:crit:xs}
  r = e^{x-s}, \hspace{1em}
  \phi = x.
\end{equation}
Observe that the scalar field $\phi$ does not depend on the scale
variable $s$ at all, and the only dependence of the metric coefficients
on the scale is through the conformal factor $e^{-2s}$. This is a
direct expression of the geometric requirement (\ref{eq:ss}) in scaling
coordinates; the homothetic Killing vector $\xi$ is simply
$-\frac{\partial\ }{\partial s}$.

\section{Gauge-Invariant Perturbations} \label{sec:pert}

To avoid complicated gauge issues of fully general perturbations, we
will use the gauge-invariant formalism developed by Gerlach and Sengupta
\cite{Gerlach&Sengupta:79,Gerlach&Sengupta:80}. This formalism describes
perturbations around a general spherically symmetric background
\begin{equation} \label{eq:pert:background}
  g_{\mu\nu} dx^\mu dx^\nu = g_{AB} dx^A dx^B
                           + r^2 \gamma_{ab} dx^a dx^b,
\end{equation}
which in our case we take to be the Roberts solution (\ref{eq:metric}).
Here and later capital Latin indices take values $\{0,1\}$, and
lower-case Latin indices run over angular coordinates. $g_{AB}$ and $r$
are defined on a spacetime two-manifold, while $\gamma_{ab}$ is the
metric of the unit two-sphere.

Because the background spacetime is spherically symmetric,
perturbations around it can be decomposed in spherical harmonics.
Scalar spherical harmonics $Y_{lm}(\theta, \varphi)$ have even parity
under spatial inversion, while vector spherical harmonics
$S_{lm\,a}(\theta, \varphi) \equiv \epsilon_a^{~b} Y_{lm,b}$ have odd
parity. We will only concern ourselves with even-parity perturbations
here, since odd-parity perturbations can not couple to scalar field
perturbations. We will focus on non-spherical perturbation modes
($l \ge 1$), as the spherically symmetric case ($l=0$) was studied
earlier \cite{Frolov:97}. For clarity, angular indices $l, m$ and the
summation over all harmonics will be suppressed from now on. The most
general even-parity metric perturbation is
\begin{eqnarray} \label{eq:pert:metric}
  \delta g_{\mu\nu} dx^\mu dx^\nu &=& h_{AB} Y dx^A dx^B \nonumber\\
    && + h_A Y_{,b} (dx^A dx^b + dx^b dx^A) \nonumber\\
    && +r^2 [K Y \gamma_{ab} + G Y_{:ab}] dx^a dx^b,
\end{eqnarray}
and the scalar field perturbation is
\begin{equation} \label{eq:pert:field}
  \delta \phi = F Y.
\end{equation}
As you can see, metric perturbations are described by a two-tensor
$h_{AB}$, a two-vector $h_A$, and two two-scalars $K$ and $G$; the
scalar field perturbation is described by a two-scalar $F$. However,
these perturbation amplitudes do not have direct physical meaning, as
they change under the (even-parity) gauge transformation induced by the
infinitesimal vector field
\begin{equation} \label{eq:pert:gauge}
  \xi_\mu dx^\mu = \xi_A Y dx^A + \xi Y_{,a} dx^a.
\end{equation}
One can construct two gauge-invariant quantities from the metric
perturbations
\begin{eqnarray} \label{eq:pert:gi:metric}
  k_{AB} &=& h_{AB} - 2 p_{(A|B)},\nonumber\\
  k &=& K - 2 v^A p_A,
\end{eqnarray}
and one from the scalar field perturbation
\begin{equation} \label{eq:pert:gi:field}
  f = F + \phi^{,A} p_A,
\end{equation}
where
\begin{equation}
  v_A = \frac{r_{,A}}{r}, \hspace{1em}
  p_A = h_A - \frac{r^2}{2}\, G_{,A}.
\end{equation}
Only gauge-invariant quantities have physical meaning in the
perturbation problem. All physics of the problem, including the
equations of motion and boundary conditions, should be written in terms
of these gauge-invariant quantities. Once gauge-invariant quantities
have been identified, one is free to convert between gauge-invariant
perturbation amplitudes and their values in whatever gauge choice one
desires.

We will work in longitudinal gauge ($h_A = G = 0$), which is
particularly convenient since perturbation amplitudes in it are just
equal to the corresponding gauge-invariant quantities. The above
condition fixes the gauge uniquely for non-spherical modes. (There is
some gauge freedom left over for the $l=0$ mode, but remember that we
are only concerned with higher $l$ modes.) Expressions for the
components of the linear perturbation equations
\begin{eqnarray}
  &\delta R_{\mu\nu} = 4 \phi_{(,\mu} \delta \phi_{,\nu)}&, \\
  &\delta (\Box\phi) = 0&
\end{eqnarray}
for a fully general perturbation in longitudinal gauge are collected in
Appendix~\ref{sec:long}. By inspection of the $\theta\varphi$ component
of the equations, it is clear that the equations of motion require that
$h_{uv}=0$ for $l \ge 1$. With the change of notation $h_{uu}=U$ and
$h_{vv}=V$, the remaining equations of motion for non-spherical modes
are
\widetext
\begin{mathletters} \label{eq:pert}
\begin{equation} \label{eq:pert:box}
  4 (u^2-uv) F_{,vu} - u U_{,v} - u K_{,u} + v K_{,v} + v V_{,u}
  - 2 u F_{,u} + 2 (2u-v) F_{,v} + 2 l(l+1) F = 0,
\end{equation}
\begin{equation} \label{eq:pert:r:uu}
  - 2 (u^2-uv) K_{,uu} + u U_{,u} + (2u-v) (U_{,v} - 2 K_{,u})
  - 4 v F_{,u} + l(l+1) U = 0,
\end{equation}
\begin{equation} \label{eq:pert:r:uv}
  - (u^2-uv) (U_{,vv} + 2 K_{,vu} + V_{,uu})
  + u U_{,v} + u K_{,u} - (2u-v) (K_{,v} + V_{,u})
  + 2 u F_{,u} - 2 v F_{,v} = 0,
\end{equation}
\begin{equation} \label{eq:pert:r:vv}
  - 2 (u^2-uv) K_{,vv} + 2 u K_{,v}
  - u V_{,u} - (2u-v) V_{,v} + 4 u F_{,v} + l(l+1) V = 0,
\end{equation}
\begin{equation} \label{eq:pert:r:tt}
  2 (u^2-uv) K_{,vu} - u U_{,v} - 2 u K_{,u} + (2u-v) (2 K_{,v} + V_{,u})
  - 2 K + l(l+1) K + 2 V = 0,
\end{equation}
\begin{equation} \label{eq:pert:r:ut}
  (u^2-uv) (U_{,v} + K_{,u}) + 2 v F = 0,
\end{equation}
\begin{equation} \label{eq:pert:r:vt}
  (u-v) (V_{,u} + K_{,v}) - 2 F = 0.
\end{equation}
\end{mathletters}
\narrowtext
Equation (\ref{eq:pert:box}) comes from the scalar wave equation, and
equations (\ref{eq:pert:r:uu}--\ref{eq:pert:r:vt}) are the $uu$, $uv$,
$vv$, $\theta\theta$, $u\theta$, and $v\theta$ components of the
Einstein equations, correspondingly. As usual with a scalar field, the
system (\ref{eq:pert}) has one redundant equation, so equation
(\ref{eq:pert:r:uv}) is satisfied automatically by virtue of other
equations. Equations (\ref{eq:pert:r:ut}) and (\ref{eq:pert:r:vt}) are
constraints, and the remaining four equations are dynamic equations for
four perturbation amplitudes $U$, $V$, $K$, and $F$.

Boundary conditions for the system (\ref{eq:pert}) are specified at
$v=0$ and the spatial infinity. Continuity of matching with flat
spacetime at the hypersurface $v=0$ requires the vanishing of the
perturbations there. We also require well-behavedness of the
perturbations at ${\cal I}^-$ and ${\cal I}^+$, so that the
perturbation expansion holds. Thus, the boundary conditions are
\begin{eqnarray} \label{eq:pert:bc}
  &U=V=K=F=0 \text{ at } v=0,& \nonumber\\
  &U, V, K, F \text{ are bounded at } u=-\infty \text{ and } v=+\infty.&
\end{eqnarray}
Equations (\ref{eq:pert}) together with boundary conditions
(\ref{eq:pert:bc}) constitute our eigenvalue problem.

\section{Decoupling of Perturbation Equations} \label{sec:decouple}

It is possible to decouple the dynamic equations
(\ref{eq:pert:box}--\ref{eq:pert:r:tt}) by combining them with the
constraints (\ref{eq:pert:r:ut}) and (\ref{eq:pert:r:vt}), and their
first derivatives. After somewhat cumbersome algebraic manipulations,
which we will not show here, the system of linear perturbation
equations (\ref{eq:pert}) can be rewritten as
\widetext
\begin{mathletters} \label{eq:uv}
\begin{equation} \label{eq:uv:f}
  2 (u^2-uv) F_{,vu} - u F_{,u} + (2u-v) F_{,v}
  + \frac{2vF}{u-v} + l(l+1) F = 0,
\end{equation}
\begin{equation} \label{eq:uv:u}
  2 (u^2-uv) U_{,vu} + u U_{,u} + 3 (2u-v) U_{,v} + l(l+1) U = 0,
\end{equation}
\begin{equation} \label{eq:uv:v}
  2 (u^2-uv) V_{,vu} - 3 u V_{,u} - (2u-v) V_{,v} + l(l+1) V = 0,
\end{equation}
\begin{equation} \label{eq:uv:k}
  2 (u^2-uv) K_{,vu} - u K_{,u} + (2u-v) K_{,v} - 2 K + l(l+1) K =
  - 2 V - \frac{4uF}{u-v},
\end{equation}
\begin{equation} \label{eq:uv:ut}
  u U_{,v} + u K_{,u} + \frac{2vF}{u-v} = 0,
\end{equation}
\begin{equation} \label{eq:uv:vt}
  V_{,u} + K_{,v} - \frac{2F}{u-v} = 0.
\end{equation}
\end{mathletters}
This decoupled system of partial differential equations can be further
simplified by exploiting continuous self-similarity of the background
to separate spatial and scale variables. With this intent, we rewrite
equations (\ref{eq:uv}) in terms of the scaling coordinates
(\ref{eq:coord:xs})
\begin{mathletters} \label{eq:xs}
\begin{equation} \label{eq:xs:f}
  \frac{1}{2}\, (1-e^{-2x}) F_{,xx} + F_{,xs} + F_{,s} - 2 (1-e^{-2x}) F + l(l+1) F = 0,
\end{equation}
\begin{equation} \label{eq:xs:u}
  \frac{1}{2}\, (1-e^{-2x}) U_{,xx} + U_{,xs} - 2 U_{,x} - U_{,s} + l(l+1) U = 0,
\end{equation}
\begin{equation} \label{eq:xs:v}
  \frac{1}{2}\, (1-e^{-2x}) V_{,xx} + V_{,xs}+ 2 V_{,x} + 3 V_{,s} + l(l+1) V = 0,
\end{equation}
\begin{equation} \label{eq:xs:k}
  \frac{1}{2}\, (1-e^{-2x}) K_{,xx} + K_{,xs} + K_{,s} - 2 K + l(l+1) K = - 2 V - 4 e^{-2x} F,
\end{equation}
\begin{equation} \label{eq:xs:ut}
  U_{,x} - (1-e^{2x}) K_{,x} + 2 e^{2x} K_{,s} - 4 (1-e^{2x}) F = 0,
\end{equation}
\begin{equation} \label{eq:xs:vt}
  K_{,x} - (1-e^{2x}) V_{,x} + 2 e^{2x} V_{,s} + 4 F = 0.
\end{equation}
\end{mathletters}
\narrowtext
We decompose the perturbation amplitudes into modes that grow
exponentially with the scale $s$ (which amounts to doing Laplace
transform on them)
\begin{eqnarray} \label{eq:laplace}
  F(x,s) &=& \sum_\kappa F_\kappa(x)\, e^{\kappa s}, \nonumber\\
  U(x,s) &=& \sum_\kappa U_\kappa(x)\, e^{\kappa s}, \nonumber\\
  V(x,s) &=& \sum_\kappa V_\kappa(x)\, e^{\kappa s}, \nonumber\\
  K(x,s) &=& \sum_\kappa K_\kappa(x)\, e^{\kappa s}.
\end{eqnarray}
The summation runs over the perturbation mode eigenvalues $\kappa$,
which could, in general, be complex. Modes with $\Re\,\kappa>0$ grow
and are relevant for critical behavior, while modes with
$\Re\,\kappa<0$ decay and are irrelevant. The growing perturbation
mode amplitudes vanish at $s=-\infty$, so the boundary condition at
${\cal I}^-$ is satisfied automatically. For clarity, the perturbation
mode subscript $\kappa$ and the explicit summation over all modes will
be suppressed from now on, so henceforth $F$, $U$, $V$, and $K$ will
mean $F_\kappa$, $U_\kappa$, $V_\kappa$, and $K_\kappa$ for the mode
with eigenvalue $\kappa$.

The decomposition (\ref{eq:laplace}) converts the system of partial
differential equations (\ref{eq:xs}) into a system of ordinary
differential equations, which is much easier to analyze:
\widetext
\begin{mathletters} \label{eq:x}
\begin{equation} \label{eq:x:f}
  \frac{1}{2}\, (1-e^{-2x}) F'' + \kappa F' + \kappa F - 2 (1-e^{-2x}) F + l(l+1) F = 0,
\end{equation}
\begin{equation} \label{eq:x:u}
  \frac{1}{2}\, (1-e^{-2x}) U'' + (\kappa-2) U' - \kappa U + l(l+1) U = 0,
\end{equation}
\begin{equation} \label{eq:x:v}
  \frac{1}{2}\, (1-e^{-2x}) V'' + (\kappa+2) V' + 3 \kappa V + l(l+1) V = 0,
\end{equation}
\begin{equation} \label{eq:x:k}
  \frac{1}{2}\, (1-e^{-2x}) K'' + \kappa K' + (\kappa-2) K + l(l+1) K = - 2 V - 4 e^{-2x} F,
\end{equation}
\begin{equation} \label{eq:x:ut}
  U' - (1-e^{2x}) K' + 2 \kappa e^{2x} K - 4 (1-e^{2x}) F = 0,
\end{equation}
\begin{equation} \label{eq:x:vt}
  K' - (1-e^{2x}) V' + 2 \kappa e^{2x} V + 4 F = 0.
\end{equation}
\end{mathletters}
The prime denotes a derivative with respect to spatial variable $x$.
These equations can be converted into standard algebraic form by the
change of variable
\begin{equation} \label{eq:coord:y}
  y = e^{2x}, \hspace{1em} x = \frac{1}{2} \ln y,
\end{equation}
so that the system (\ref{eq:x}) becomes
\begin{mathletters} \label{eq:y}
\begin{equation} \label{eq:y:f}
  y(1-y) \ddot{\Phi} + [3 - (\kappa+3)y] \dot{\Phi} - [3\kappa/2 + l(l+1)/2] \Phi = 0,
\end{equation}
\begin{equation} \label{eq:y:u}
  y(1-y) \ddot{U} + [1 - (\kappa-1)y] \dot{U} - [-\kappa/2 + l(l+1)/2] U = 0,
\end{equation}
\begin{equation} \label{eq:y:v}
  y(1-y) \ddot{V} + [1 - (\kappa+3)y] \dot{V} - [3\kappa/2 + l(l+1)/2] V = 0,
\end{equation}
\begin{equation} \label{eq:y:k}
  y(1-y) \ddot{K} + [1 - (\kappa+1)y] \dot{K} - [\kappa/2 - 1 + l(l+1)/2] K = 2 \Phi + V,
\end{equation}
\begin{equation} \label{eq:y:ut}
  \dot{U} + (y-1) \dot{K} + \kappa K + 2 y \Phi - 2 \Phi = 0,
\end{equation}
\begin{equation} \label{eq:y:vt}
  \dot{K} + (y-1) \dot{V} + \kappa V + 2 \Phi = 0.
\end{equation}
\end{mathletters}
\narrowtext
The dot denotes a derivative with respect to $y$, and we redefined the
scalar field perturbation amplitude as $F = y\Phi$ to cast the
equations into standard table form. The boundary conditions
(\ref{eq:pert:bc}) are
\begin{eqnarray} \label{eq:y:bc}
  &U=V=K=\Phi=0 \text{ at } y=1,& \nonumber\\
  &U, V, K, y\Phi \text{ are bounded at } y=+\infty.&
\end{eqnarray}
Imposed on the decoupled system of ordinary differential equations
(\ref{eq:y}), these boundary conditions give an eigenvalue problem for
the perturbation spectrum $\kappa$.

\section{Perturbation Spectrum} \label{sec:spectrum}

In the previous section we formulated an eigenvalue problem for the
spectrum of non-spherical perturbations of the critical Roberts
solution. We now proceed to solve it. Observe that equations
(\ref{eq:y:f}--\ref{eq:y:k}) governing the dynamics of the
perturbations are hypergeometric equations of the form
\begin{equation} \label{eq:hypergeom}
  y(1-y) \ddot{X} + [c - (a+b+1)y] \dot{X} - ab X = 0.
\end{equation}
Equation (\ref{eq:y:k}) is not homogeneous, but we will deal with that
shortly. The hypergeometric equation coefficients are different for
equations describing the perturbations $\Phi$, $U$, $V$, and $K$; they
are summarized in the table below.
\begin{equation} \label{eq:coeffs}
  \setlength{\arraycolsep}{1em}
  \renewcommand{\arraystretch}{1.4}
  \begin{array}{c|ccc}
    & c & a+b & ab\\ \hline
    \Phi & 3 & \kappa+2 & \frac{3}{2} \kappa + \frac{1}{2} l(l+1)\\
    U & 1 & \kappa-2 & -\frac{1}{2} \kappa + \frac{1}{2} l(l+1)\\
    V & 1 & \kappa+2 & \frac{3}{2} \kappa + \frac{1}{2} l(l+1)\\
    K & 1 & \kappa & \frac{1}{2} \kappa + \frac{1}{2} l(l+1)\\
  \end{array}
\end{equation}
Hypergeometric equations have been extensively studied; for complete
description of their properties see, for example,
\cite{Bateman&Erdelyi}. Hypergeometric equation (\ref{eq:hypergeom}) has
three singular points at $y=0,1,\infty$, and its general solution is a
linear combination of any two different solutions from the set
\begin{eqnarray} \label{eq:solns}
  X_1 &=& {\cal F}(a, b; a+b+1-c; 1-y),\nonumber\\
  X_2 &=& (1-y)^{c-a-b} {\cal F}(c-a, c-b; c+1-a-b; 1-y),\nonumber\\
  X_3 &=& (-y)^{-a} {\cal F}(a, a+1-c; a+1-b; y^{-1}),\nonumber\\
  X_4 &=& (-y)^{-b} {\cal F}(b+1-c, b; b+1-a; y^{-1}),
\end{eqnarray}
where ${\cal F}(a, b; c; y)$ is the hypergeometric function, which is
regular at $y=0$ and has ${\cal F}(a, b; c; 0) = 1$. Any three of the
functions (\ref{eq:solns}) are linearly dependent with constant
coefficients. In particular,
\begin{eqnarray} \label{eq:connection}
  X_2 &=&
    \frac{\Gamma(c+1-a-b) \Gamma(b-a)}{\Gamma(1-a) \Gamma(c-a)}\, e^{-i\pi(c-b)} X_3 +\nonumber\\
    &&\frac{\Gamma(c+1-a-b) \Gamma(a-b)}{\Gamma(1-b) \Gamma(c-b)}\, e^{-i\pi(c-a)} X_4.
\end{eqnarray}
The functions $X_1$, $X_2$ are appropriate for discussing the behavior
of solution near $y=1$, while $X_3$, $X_4$ give the behavior at infinity
\begin{eqnarray} \label{eq:asymptotics}
  X_1 = 1,\ X_2 = (1-y)^{c-a-b} &\text{ near }& y=1, \nonumber\\
  X_3 = (-y)^{-a},\ X_4 = (-y)^{-b} &\text{ near }& y=\infty.
\end{eqnarray}

As we said before, imposing the boundary conditions (\ref{eq:y:bc}) on
solutions of equations (\ref{eq:y}) leads to a perturbation spectrum.
We will now investigate what restrictions the boundary conditions place
on the hypergeometric equation coefficients. The vanishing of
perturbation amplitudes at $y=1$ rules out $X_1$ as a component of the
solution and requires that $\Re(c-a-b) > 0$ to make $X_2$ go to zero.
The solution $X_2$ has non-zero content of both $X_3$ and $X_4$ by
virtue of (\ref{eq:connection}), hence for it to be bounded at
infinity, both $\Re\,a$ and $\Re\,b$ must be positive to guarantee
convergence of $X_3$ and $X_4$. So, unless there is degeneracy, the
boundary conditions translate to the following conditions on the
hypergeometric equation coefficients:
\begin{mathletters} \label{eq:bc:coeff}
\begin{eqnarray}
  &\Re(c-a-b) > 0,& \label{eq:bc:coeff:1}\\
  &\Re\, a, \Re\, b > 0.& \label{eq:bc:coeff:infty}
\end{eqnarray}
\end{mathletters}

We are now ready to take on system (\ref{eq:y}). Take a look at
equation (\ref{eq:y:v}) for $V$. Condition (\ref{eq:bc:coeff:1}) for it
is $\Re\,\kappa<-1$, i.e. there are no growing $V$ modes! With the
amplitude of relevant $V$ perturbation modes being zero, the
constraints (\ref{eq:y:ut}) and (\ref{eq:y:vt}) become
\begin{equation} \label{eq:constraints}
  K = - \frac{\dot{U}}{\kappa}, \hspace{1em}
  \Phi = \frac{\ddot{U}}{2\kappa},
\end{equation}
and right hand side of equation (\ref{eq:y:k}) can be absorbed by the
left hand side, making the equation for $K$ homogeneous (with $c=2$).
Indeed equations (\ref{eq:y:k}) and (\ref{eq:y:f}) for $K$ and $\Phi$
are just derivatives of equation (\ref{eq:y:u}) for $U$
\begin{equation} \label{eq:master}
  y(1-y) \ddot{U} + [1 - (\kappa-1)y] \dot{U} - [-\kappa/2 + l(l+1)/2] U = 0,
\end{equation}
which is the homogeneous hypergeometric equation with coefficients
\begin{equation} \label{eq:master:coeff}
  c = 1, \hspace{1em}
  a+b = \kappa-2, \hspace{1em}
  ab = -\frac{1}{2} \kappa + \frac{1}{2} l(l+1).
\end{equation}
Imposing the boundary condition at $y=1$ for the solution of the above
equation and its derivatives, which behave like
\begin{equation}
  \left.\begin{array}{r}
    U \propto (1-y)^{3-\kappa}\\
    K \propto (1-y)^{2-\kappa}\\
    \Phi \propto (1-y)^{1-\kappa}\\
  \end{array}\right\} \text{ near } y=1,
\end{equation}
produces restriction on the non-spherical mode eigenvalue
\begin{equation}
  \Re\, \kappa < 1,
\end{equation}
which is the strongest of restrictions (\ref{eq:bc:coeff:1}) for
equations for $U$, $K$, and $\Phi$. But then
\begin{equation}
  \Re\, a + \Re\, b = \Re\, \kappa - 2 < -1,
\end{equation}
and hence $\Re\,a$ and $\Re\,b$ can not be both positive, and so the
boundary condition at infinity can not be satisfied. A more careful
investigation of degenerate cases of relation (\ref{eq:connection})
shows that the contradiction between boundary conditions at $y=1$ and
infinity still persists if $V=0$. It can only be resolved by the
trivial solution $U=K=\Phi=0$. Thus we have shown that there are no
growing non-spherical perturbation modes around the critical Roberts
solution.

In fact, an even stronger statement is true. The contradiction between
boundary conditions at $y=1$ and infinity can not be resolved by a
non-trivial solution so long as $V=0$, i.e. so long as
$\Re\,\kappa \ge -1$. Hence non-trivial non-spherical perturbation
modes of critical Roberts solution must decay faster than $e^{-s}$.

\section{Conclusion} \label{sec:conclusion}

In this paper we used the gauge-invariant perturbation formalism to
explore the critical behavior in the gravitational collapse of a
massless scalar field. Perturbing around a continuously self-similar
critical solution (the Roberts solution), we obtained an eigenvalue
problem for the spectrum of perturbations. The remarkable feature of
this model of critical scalar field collapse is that it allows an exact
analytical treatment of the perturbations as well as of the critical
solution, due to the highly symmetric background.

An exact analysis of the perturbation eigenvalue problem reveals that
there are no growing non-spherical perturbation modes. However, there
are growing spherical perturbation modes. Their spectrum is continuous
and occupies a big chunk of the complex plane \cite{Frolov:97}. In view
of these findings, the following picture of dynamics of scalar field
evolution near self-similarity emerges: As we evolve generic initial
data which is sufficiently close to the critical Roberts solution,
non-spherical modes decay and the solution approaches the spherically
symmetric one. Asymmetry of the initial data does not play a role in
the collapse. The growing spherical modes, on the other hand, drive the
solution farther away from the continuously self-similar one. In this
sense, the critical Roberts solution is an intermediate attractor for
non-spherical initial data.

An interesting question, which is not answered by perturbative
calculations, is the further fate of the scalar field evolution as it
gets away from the Roberts solution. In all likelihood, it evolves
towards the discretely self-similar Choptuik solution, which is a local
attractor of lower codimension (one), as the continuous self-similarity
is broken by oscillatory growing modes. After staying near the Choptuik
solution for a while, the scalar field will eventually disperse or
settle into a black hole, with these final states being global
attractors in the phase space of field configurations. This evolution
from attractor to attractor in phase space is somewhat analogues to a
ball rolling down the stairs, going from a step to a lower step, until
it reaches the bottom.

The results of this paper shed some light at the complicated problem of
critical collapse of generic initial data from the analytical
viewpoint, confirming the hypothesis that critical phenomena are not
restricted to spherical symmetry. Investigation of the fate of a scalar
field solution as it breaks away from continuous self-similarity, as
outlined above, will further our understanding of the dynamics of
scalar field collapse, and presents an interesting (and challenging)
analytical problem. Numerical simulations might also help to establish
a clearer picture of near-critical scalar field evolution.

\section*{Acknowledgments}

This research was supported by the Natural Sciences and Engineering
Research Council of Canada and by the Killam Trust.

\appendix
\section{Perturbations in Longitudinal Gauge} \label{sec:long}

In this appendix we collect expressions for components of the perturbed
Einstein-scalar equations calculated in longitudinal gauge ($h_A=G=0$).
The perturbed metric in longitudinal gauge is
\begin{equation}
  ds^2 =
    h_{uu} Y du^2 - 2(1 - h_{uv} Y) du\, dv + h_{vv} Y dv^2
    + (1 + K Y) r^2 d\Omega^2,
\end{equation}
and the perturbed scalar field is
\begin{equation}
\phi = \frac{1}{2} \left[1 - \frac{v}{u}\right] + F Y.
\end{equation}
The Einstein equations for scalar field are equivalent to the
vanishing of the tensor $E_{\mu\nu} = R_{\mu\nu} - 2 \phi_{,\mu} \phi_{,\nu}$.
Its non-trivial components, calculated to the first order in the
perturbation amplitude using the above metric and scalar field, are
\begin{mathletters}
\begin{eqnarray}
&& E_{uu} =
  \frac{1}{2} \Big[
    - 2 (u^2-uv) K_{,uu} + u h_{uu,u} \nonumber\\ && \mbox{\hspace{4.3em}}
    + (2u-v) (h_{uu,v} - 2 h_{uv,u} - 2 K_{,u}) \nonumber\\ && \mbox{\hspace{4.3em}}
    - 4 v F_{,u} + l(l+1) h_{uu}
  \Big]\, \frac{Y}{u^2-uv},
\end{eqnarray}
\begin{eqnarray}
&& E_{uv} = 
  - \frac{1}{2} \Big[
    (u^2-uv) (h_{uu,vv} - 2 h_{uv,vu} + h_{vv,uu} + 2 K_{,vu}) \nonumber\\ && \mbox{\hspace{4.3em}}
    - u h_{uu,v} + (2u-v) (h_{vv,u} + K_{,v}) - u K_{,u} \nonumber\\ && \mbox{\hspace{4.3em}}
    + 2 v F_{,v} - 2 u F_{,u} - l(l+1) h_{uv}
  \Big]\, \frac{Y}{u^2-uv},
\end{eqnarray}
\begin{eqnarray}
&& E_{vv} = 
  \frac{1}{2} \Big[
    - 2 (u^2-uv) K_{,vv} + 2 u h_{uv,v} \nonumber\\ && \mbox{\hspace{4.3em}}
    - u h_{vv,u} - (2u-v) h_{vv,v} + 2 u K_{,v} \nonumber\\ && \mbox{\hspace{4.3em}}
    + 4 u F_{,v}
    + l(l+1) h_{vv}
  \Big]\, \frac{Y}{u^2-uv},
\end{eqnarray}
\begin{eqnarray}
&& E_{u\theta} = 
  - \frac{1}{2} \Big[
    (u^2-uv) (h_{uu,v} - h_{uv,u} + K_{,u}) \nonumber\\ && \mbox{\hspace{4.3em}}
    + (2u-v) h_{uv} + 2 v F
  \Big]\, \frac{Y_{,\theta}}{u^2-uv},
\end{eqnarray}
\begin{eqnarray}
&& E_{u\varphi} = 
  - \frac{1}{2} \Big[
    (u^2-uv) (h_{uu,v} - h_{uv,u} + K_{,u}) \nonumber\\ && \mbox{\hspace{4.3em}}
    + (2u-v) h_{uv} + 2 v F
  \Big]\, \frac{Y_{,\varphi}}{u^2-uv},
\end{eqnarray}
\begin{eqnarray}
&& E_{v\theta} = 
  - \frac{1}{2} \Big[
    (u-v) (- h_{uv,v} + h_{vv,u} + K_{,v}) \nonumber\\ && \mbox{\hspace{4.3em}}
    - h_{uv} - 2 F
  \Big]\, \frac{Y_{,\theta}}{u-v},
\end{eqnarray}
\begin{eqnarray}
&& E_{v\varphi} = 
  - \frac{1}{2} \Big[
    (u-v) (- h_{uv,v} + h_{vv,u} + K_{,v}) \nonumber\\ && \mbox{\hspace{4.3em}}
    - h_{uv} - 2 F
  \Big]\, \frac{Y_{,\varphi}}{u-v},
\end{eqnarray}
\begin{eqnarray}
&& E_{\theta\theta} = 
  \frac{1}{2} \Big[
    2 (u^2-uv) K_{,vu} - u h_{uu,v} \nonumber\\ && \mbox{\hspace{4.3em}}
    + (2u-v) (h_{vv,u} + 2 K_{,v}) - 2 u K_{,u} \nonumber\\ && \mbox{\hspace{4.3em}}
    - 2 h_{uv} + 2 h_{vv} - 2 K + l(l+1) K
  \Big]\, Y \nonumber\\ && \mbox{\hspace{3em}}
  + h_{uv} Y_{,\theta\theta},
\end{eqnarray}
\begin{equation}
E_{\varphi\varphi} = \sin^2\theta\, E_{\theta\theta},
\end{equation}
\begin{equation}
E_{\theta\varphi} =
  h_{uv} (Y_{,\theta\varphi} - \cot\theta\, Y_{,\varphi}).
\end{equation}
\end{mathletters}
The scalar field equation requires the vanishing of $\Box\phi$, which,
calculated to first order using the above metric and scalar field, is
\begin{eqnarray}
&& \Box\phi = 
  \frac{1}{2} \Big[
    - 4 (u^2-uv) F_{,vu} + u h_{uu,v} - v h_{vv,u} \nonumber\\ && \mbox{\hspace{4.3em}}
    + u K_{,u} - v K_{,v} + 2 u F_{,u} - 2 (2u-v) F_{,v} \nonumber\\ && \mbox{\hspace{4.3em}}
    - 2 l(l+1) F
  \Big]\, \frac{Y}{u^2-uv}.
\end{eqnarray}

%%% Bibliography

%\bibliographystyle{prsty}	% Custom bibliography style
%\bibliography{../journals,../cosmology,../critical,../frolov,paper}

%%% Figures

\end{document}